# Microscopic Aspects of Stretched Exponential Relaxation (SER)

J. C. Phillips

Dept. of Physics and Astronomy, Rutgers University, Piscataway, N. J., 08854-8019

## Abstract

The Scher-Lax-Phillips (SLP) universal minimalist model quantitatively explains stretching fractions $\beta(T_g)$ for a wide variety of relaxation experiments (nearly 50 altogether) on electronic and molecular glasses and deeply supercooled liquids by assuming that quasi-particle excitations indexed by Breit-Wigner channels diffuse to traps (sinks). This model is effective here in discussing in detail three experiments: luminescence in isoelectronic Zn(Se,Te) alloys, fibrous relaxation in orthoterphenyl (OTP) and related glasses and melts up to $1.15T_g$, and relaxation of binary chalcogen melts probed by spin-polarized neutrons (T as high as $1.5T_g$). The model is also compared to several other recent theories.

**1. Introduction**

Because the theory of SER is still a work in progress, the phenomenon itself can be said to be the oldest unsolved problem in science. Many electrical and optical phenomena exhibit SER with probe relaxation $I(t) \sim \exp[-(t/\tau)^\beta]$, with $0 < \beta < 1$, but in recent decades the most reliable data have been obtained on atomic and molecular relaxation of glasses and deeply supercooled liquids. As the data base grew, the need for a quantitative theory increased; this need was finally met by the SLP diffusion-to-traps model [1,2], which yields a remarkably simple expression for $\beta$, given by $d^*/(d^* + 2)$. This suggests that the low-temperature limit for $\beta$ could be 3/5 (at high T simple exponential relaxation usually occurs, $\beta = 1$), providing that the dimensionality of the configuration space in which the relaxation occurs is simply $d^* = d = 3$. Other values of $\beta$ imply more complex configuration spaces, which are likely to occur in glasses and deeply supercooled liquids, with their exponentially large viscosities. Surprisingly enough, there are many glasses which do exhibit $\beta(T_g) = 3/5$ (within a few %).

While the complexity of glass-forming materials and the large data bases required to explore SER discouraged research on this problem for many years, computers with large memories have made the problems much more accessible, although the results are still very sensitive to experimental design. This paper discusses three very well-designed experiments which shed light on the traps to which excitations diffuse, and the configuration space in which the diffusion takes place. The discussion takes place within the general framework of theories that analyze how specific kinds of chemical bonding lead to efficient space-filling without crystallization, but it also requires much detailed information on specific glassy materials.

The best glass formers (slowest cooling rates, most homogeneous materials) are found in oxide network glass alloys, followed by chalcogenide alloy network glasses, but many molecular glasses are known, together with polymers and even metallic and colloidal glasses. The characteristic feature of good glass formers is that they efficiently fill space, as reflected by constraint theory for oxide and chalcogenide network glass alloys (non-central forces) [3,4], and by free volume theories for metallic glasses (central forces) [5,6]. Constraint theory of hydrogen bonding also accurately describes the glass-forming tendencies of small molecular alcohols and saccharides [7]. The microscopic structure of good glasses often persists through the glass transition at $T = T_g$, up to a crossover temperature $T_0 \sim (T_g + T_m)/2$ or larger, as reflected in the viscosity (fragility) of supercooled liquids [8]. Thus one can suppose that this structure consists of clusters percolatively connected in the glass, with the rigid percolative paths broken in the deeply supercooled liquid, and the clusters themselves decomposing only for $T > T_0$. Rigidity percolation is the characteristic feature of network glasses [9,10].

**2. Topological and Geometrical Models of SER**

Generally relaxation in a glass or deeply supercooled liquid is a complex process, similar in some respects to capture of slow neutrons by nuclei. The standard Breit-Wigner method for treating the latter problem introduces separate and distinct reaction channels [11], and glassy relaxation can be similarly discussed using relaxation channels. These are supposed to represent non-vibrational degrees of freedom which can lead to relaxation. As these are not known in detail, it was long believed that SER cannot be treated quantitatively. However, in the 1970-1980's it was realized [12,13] that the central feature of SER, the stretching fraction $\beta$, must be a topological parameter, dependent only on the effective dimensionality $d^*$ of the configuration space in which the excitation diffuses, and given explicitly by $\beta(T_g) = d^*/(d^* + 2)$. This apparently only replaces one unknown dimensionless parameter ($\beta$) with another unknown dimensionless parameter ($d^*$). By the 1990's the early data base (which was littered with artifacts, based on partially crystallized and/or inhomogeneously hydrated polymers, etc.) had been superceded by many excellent studies of carefully prepared glasses and supercooled liquids, as well as numerical simulations with supercomputers. When these data were collected, they exhibited simple quantitative patterns for $\beta(T_g)$, with easily justified values of $d^*$ [1,2].

At this point one can pause to take stock. Because of the exponential complexity of glasses, as reflected by superexponential increases in viscosity $\eta(T)$ as $T \rightarrow T_g$, conventional polynomial methods based on Hamiltonians, partition functions, or various random but still restricted models (including lattice percolation models) have never reliably derived $\beta(T_g)$. Topology, however, transcends the restrictions of polynomial models. Moreover, the derivations [12,13,1] of $\beta(T_g) = d^*/(d^* + 2)$ rely only on the diffusion equation and the existence of randomly distributed trapping sites at which excitation energies are dissipated, so that the trap model is truly minimal.

How successful is this minimal model? Results [1,2] from large-scale numerical simulations of simple models, experimental values for network glasses, polymers, ionic fused salts and organic molecular glasses are listed in Table 1, and it is clear by any standard that the model is very successful indeed, in fact, nearly universal. It is amusing to note that the conclusion to [1] suggested that "in ten years the comparison between theory and experiment [may be] even closer than [now]", and that this is exactly what happened [2].

Whether or not such trapping sites could exist in glasses and deeply supercooled liquids was uncertain until the 1990's, when the successes of both simulations and a variety of experiments left no doubt that the relation $\beta(T_g) = d^*/(d^* + 2)$ is often very accurate. Thus it is arguable that these successes by themselves provide convincing circumstantial evidence for the existence of discrete traps with abrupt boundaries, outside of which the traditional diffusion equation is valid for relaxing excitations. This conclusion is not obvious, as the glassy or deeply supercooled liquid matrix outside the rigid traps may not have the properties of a normal liquid: for example, it has been claimed that near $T_g$ the Stokes-Einstein relation between viscosity and self-diffusion coefficient fails in fragile molecular glasses such as OTP [14], as discussed in Sec. 4. In one very simple and direct case, the diffusion of Ag markers in a chalcogenide alloy glass, the traps have been identified unambiguously in realistic numerical simulations [15], thus putting to rest any doubts concerning the need for microscopic justification of the trap model.

Viscosity and self-diffusion decoupling raises questions about the validity of the universal minimalist trap model, and has led to efforts to derive SER in other ways based on particle scattering theory [16]. Unfortunately, these efforts failed to derive the key result of the trap model, $\beta(T_g) = d^*/(d^* + 2)$. Examining their analysis, we find that the discussed liquid correlation functions are *equilibrium* functions that do not distinguish between scattering and relaxation. This distinction is important ergodically: in the trap model, excitation energies are dissipated in the rigid traps, never (in the glass) or seldom (in deeply supercooled liquids) to return to the liquid matrix (eventually they reach the sample boundaries via chains of trap states [17]). In deeply supercooled liquids the observed rapidly diffusing states are analogous to the fast ion states of solid electrolytes, which are known to follow percolative paths composed of structural units with free volumes *F* above average [18,19]. The paths are smoothly self-organized to minimize free volume differences δ*F* between successive filamentary cells, which minimizes scattering due to density fluctuations. Large scale numerical simulations then predict ionic conductivities over 11 orders of magnitude, and show that the activation energy $E_\sigma$ for ionic conduction (analogous to a pseudogap or superconductive gap energy) scales with $F^{d/3}$ with $d = 1$.

It is instructive to compare the topological $\beta(T_g) = d^*/(d^* + 2)$ relation with the results of a recent simulation using an optimized hopping (minimax) geometrical trap model for electronic relaxation [20,21]. In the geometrical model there are four non-observable parameters, the densities of transport and trapping sites, $N_o$ and $N_T$, and the radii of these sites, a and R. The model does not give SER asymptotically at large times, but it does give curves that are accurately fitted by SER at short and intermediate times. The curves depend on two unknown parameters, $N_o/N_T$ and a/R, as do the fitted values of $\beta$, which is disappointing compared to the topological result $\beta(T_g) = d^*/(d^* + 2)$. Even in the simplest possible case, strongly disordered Zn(Se,Te) alloys, the meaning of $N_o/N_T$ and a/R is elusive, while d* is easily interpreted (Sec. 3).

The most subtle aspect of the SLP universal minimalist model is the definition of $d^* = fd$. Here d is the dimension of Cartesian scattering space, and f measures the ratio of the numbers of effective relaxation channels to their total number. In many simple cases involving short-range forces only (sphere mixtures, metallic glasses, fully cross-linked network glasses) f = 1, leading to $\beta = 3/5$. However, in more complex cases (for example, Se chains, polymers, fused salts, many electronic glasses) f = ½, so $\beta = 3/7$. The pervasive pattern here is that in the simultaneous presence of short- and long-range forces there are two equally weighted sets of scattering channels, only one of which is effective for relaxation. This pattern is apparent in Table I, and phenomenologically speaking it is unambiguous.

Long-range forces can arise from stress (Se and polymers) or from Coulomb forces (fused salts, many electronic glasses). Recently a very elegant case (well-controlled pseudo-binary semiconductors) has been studied, time-resolved luminescence from Zn(Se,Te) alloys (useful for orange light-emitting diodes), that shows the crossover from non-diffusive Debye relaxation ($\beta = 1$), first to maximally localized relaxation with $\beta = 3/5$, and then to the lower bound $\beta = 3/7$ (Sec. 3). A few special cases are also known where $f = 1/3$ and $d^* = 1$; of course, these must occur because of special geometries (Sec. 4). The equal weighting of short-and long-range channels is plausible in two respects: by analogy with the Ewald method for optimizing Coulomb lattice sums, and by realizing that in the nearly steady state as soon as the two channels become unbalanced, the more numerous one will lose its extra weight by relaxing faster (a kind of competitive detailed balance).

The question of determining $d^*$ is relatively simple so long as relaxation is dominated by quasiparticles, but glasses and deeply supercooled liquids often contain clusters, which can be regarded as part of optimized space-filling (medium-range precursors of long-range crystalline order). In Q-dependent diffraction experiments on molecular glasses the quasi-particles are identified with momentum transfers $Q = Q_1$, where the largest peak in the scattering function $S(Q)$ occurs at $Q = Q_1$. In many molecular glasses and deeply supercooled liquids a weaker but still narrow peak (often called the first sharp diffraction peak, or FSDP, or Boson peak in Raman scattering) occurs at $Q = Q_B \sim 0.4Q_1$. The Boson peak is associated with extended topological defects (clusters) in space-filling glasses and deeply supercooled liquids, with values of $\beta(Q_B,T_g)$ qualitatively different from those of $\beta(Q_1,T_g)$. Sec. 5 extends SLP theory to explain SER trends and magnitudes for the Boson $Q = Q_B$ peak in deeply supercooled chalcogenide liquid alloys.

### 3. Luminescence in Isoelectronic Zn(Se,Te) Alloys

Probably the most reliable early measurements of SER in luminescence were the pump/probe photoinduced absorption studies of Cd(S,Se) nanocrystallites used in commercial optical filters [22]; similar results were obtained in fullerene $C_{60}$ films [1,2]; in both cases $\beta(T)$ leveled off at low T near 0.40. Later further luminescence studies on porous Si showed $\beta(T)$ again leveled off at low T near 0.40 [2]. These optical experiments on quite diverse samples strongly support the SLP universal minimalist model of SER [1,2] for the case of mixed short- and long-range interactions ($d^* = 3/2$).

In each case one could argue that Coulomb blockade is involved in the low-T stretching. Si nanodots with diameters small compared to exciton radii gave d* = 3, as expected (short-range forces only).

These older data are now dramatically confirmed by modern time-resolved studies of isoelectronic luminescence from ZnSe$_{1-x}$Te$_x$ alloys [23], with results for $\tau(x)$ and $\beta(x)$ shown in Fig. 1. Although the energy gaps at x = 0 and x = 1 are nearly equal, this gap is strongly bowed downward, and the band edge shifts from 2.7 eV (x = 0) to 2.05 eV for x = 0.35. Because hole masses are much larger than electron masses, as x increases from 0, valence band edge offsets associated with Te sites localize holes on Te clusters, which then bind electrons, leading to exciton-mediated recombination. [24] observed a continuous Localized – Extended (L-E) transition from the recombination through free and bound exciton states to the recombination of excitons localized by the compositional fluctuations of the mixed crystal in the concentration region of about x = 0.25.

[23] observed that the narrow coherent exciton recombination band found at x = 0 with $\beta(0)$ = 1 was already broadened at x = 0.005, and the exciton lifetime increased by a factor of order 100 between x = 0 and x = 0.08 (longest lifetime, most localized exciton). Meanwhile, $\beta(x)$ dropped rapidly along an S-like curve centered on x = 0.05, to reach $\beta(0.10)$ = 3/5, which is *exactly* the value expected for a fully localized (longest lifetime) state whose kinetics are dominated by short-range interactions. At x = 0.10, $\beta(x)$ abruptly changes slope, decreasing more slowly to a broad, rather flat minimum with $\beta(0.22)$ = 0.45, which is very close to the value 3/7 expected for a perfectly balanced mixture of short- and long-range interactions. According to [24], x = 0.20 - 0.25 is where the L-E transition takes place, effectively mixing short- and long-range forces, so again the 1996 SLP universal minimalist theory has *quantitatively* explained all the key features of this elegant 2008 experiment, as indicated in Fig. 1. Of course, at other values of x than 0.10 and 0.22, a mixture of mechanisms will determine $\beta(x)$. Thus at x = 0 or 1, there is no diffusion, and there is simply exponential recombination with $\beta$ = 1. Between 0 and 0.10, one has a mixture of free and bound exciton recombination, and so on. It is noteworthy

that the intuitively plausible geometrical trap model [20] is dependent on non-topological parameters, and therefore is unable to identify the key features of Fig. 1.

Looking at Fig. 1, many scientists would readily concede that there are two special values of x that correspond to $\beta$ = 3/5 and 3/7, but they would still argue that for almost all values of x these special values are mixed with each other and with $\beta$ = 1. However, it is important here to remember that $ZnSe_{1-x}Te_x$ alloys by themselves are not glasses. For most values of x the glassy behavior of luminescence in these alloys involves mixing the $\beta$ = 1, 3/5 and 3/7 channels because the disorder of the alloys still takes place in a crystalline framework. The internal ordering of the alloys involves strain energies that are dependent on many bonding energies other than those involved in band-edge exciton localization and recombination. Only at the analytic extrema (maximum $\tau$ or minimum $\beta$) do we obtain single-channel behavior. What is remarkable about Fig. 1 is that in spite of these additional factors, the relaxation still follows the Kohlrausch form, and merely mixes the three $\beta$ channels to generate an intermediate value of $\beta$. In real glasses, as shown in Table 1, the relaxation is usually dominated by a single channel with $\beta$ = 3/5 or 3/7. Other values of $\beta$ do occur occasionally, corresponding to mixed channels, examples of which are discussed in [1,2]. Another such example will be discussed in Sec. 5; once the "pure" cases are understood, it becomes relatively easy to explain the mixed cases.

4. **Orthoterphenyl Revisited**

Because it is commercially available in high purity, and because of its symmetrical three-ring planar structure, resembling anthracene, OTP is the most studied organic molecular glass former, apart from simple alcohols and sugars [7]. Already in Table 1 there are four entries for OTP, and five values for $\beta$, three close to 3/5 and two close to 3/7 (the specific choices depend on probe and (in the case of multi-dimensional NMR), pump history as well (see [1,2] for details, which are important, and which also challenge all other non-topological theories).

Although dielectric values of $\beta$ (based on plane parallel capacitance configurations) were included in [1,2], it was noted there that these seldom agree with the values measured by other methods. The discordances arise from the combination of extrinsic factors (dielectric relaxation is a simple measurement, hence it is often made on poor samples, and properly curve-fitting even good data requires high numerical accuracy) and a fundamental intrinsic factor (forced relaxation in an electric field is different from free relaxation) [25]. Often NMR values of β also disagree with the values measured by other methods. This happens because NMR is extremely sensitive to small concentrations of clusters that produce narrow resonance lines.

Optical measurements using polarized light to photoselect an orientationally anisotropic subset of probe molecules by photbleaching can also emphasize clusters. [26] studied SER in OTP in this way and found near $T_g$ that β(OTP) = 0.34 and β(anthracene in OTP) = 0.39. Values of β for five other probe molecules were also obtained, all of which satisfied β ≥ 0.60. These data have a simple explanation in terms of the SLP model (see Fig. 2). There are two branches to the measured values of β, plotted in Fig. 2 as a function of the probe hydrodynamic volume normalized to the OTP host volume. In OTP itself the photoselected clusters are fiber-like stacks of OTP, with fluctuations in the stacks of each molecular planar axis around the local azimuthal angle, with d* = 1. Anthracene can be fitted well enough into such a glassy stack, which is why its β is close to that of OTP itself. The other molecules do not fit into the fibrous stacks, and they relax independently of the OTP fibers. Starting with tetracene (β = 0.60(6)), β increases smoothly with probe volume. This behavior is exactly what one would expect, as the volume of tetracene is close to that of OTP, while it obviously will not fit into three-ring stacks, so it should have d* = 3 and β = d*/(d* + 2) = 0.60. As the probe volume increases, the number of degrees of freedom available for relaxation by collisions with the smaller host OTP increases, and β increases smoothly, as reported. SLP theory is exact for all these OTP data; SER in OTP has recently been discussed inconclusively [16].

Of course, at the time the data were reported (1995) there was no SLP model, and had such a model been available, it is unlikely that it would have been taken seriously. Even so, nowadays perhaps it should be, because there is now excellent evidence [27,28] for the existence of such molecular fibers in OTP. First, at $T_g$ + 3K OTP exhibits anomalously enhanced (by a factor of 100) translational diffusion, as measured by surface desorption. Structurally this is most easily explained by capillary diffusion along fibrous surfaces. Note that such capillary diffusion will not invalidate the main assumption of the SLP trap model, that the bulk diffusion outside the traps is normal, because the fibers themselves are the traps.

Self-diffusion controls crystallization in OTP for most of the supercooled liquid regime, but at temperatures below $T_g + 10$ K, the reported crystallization rate increases suddenly while the self-diffusion coefficient does not. This regime ("diffusionless crystallization", DC) has been observed in other molecular glass formers, notably ROY, currently the top system for the number (seven) of coexisting polymorphs of known structures. Some polymorphs did not show DC growth, while others did; the polymorphs showing DC growth changed growth morphologies with temperature, from faceted single crystals near the melting points, to fiber-like crystals near $T_g$. The DC mode was disrupted by the onset of the liquid's structural relaxation but could persist well above $T_g$ (up to 1.15 $T_g$) in the form of fast-growing fibers [28]. The SLP theory predicts that near $T_g$ $\beta$(DC mode liquids) ~ 0.35, while $\beta$(not DC mode liquids) ~ 0.6. This is a strong prediction for the seven coexisting polymorphs of the ROY system, and if confirmed, it would establish a more detailed topological criterion for DC growth than is provided by current free volume models [28].

The reader will have noticed that (including the values in Table I) we have seven different values for $\beta$(OTP), including the two branches of Fig. 2. These occur as 3/5 (four times), 3/7 (twice) and 1/3 (once). It is impossible to explain these multiple values (much less their narrowly clustering around the same "magic" fractions that occur for many other materials!) by using any geometrical model based only on free volume concepts and spatial heterogeneity. In particular, the attempt to extend the Gibbs equilibrium nucleation model (a spherical model, independent of dimensionality) to discuss SER by using modified Gaussian fluctuations [29], while qualitatively interesting, is fundamentally unsound and in fact misses the essential topologically asymptotic nature of SER in much the same way as the spherical geometrical model [20]. [29] is inadequate in many different ways, but the notion that the glass transition is somehow governed by "random" (Gaussian) statistics is perhaps its most unsatisfactory feature.

## 5. Rigid Cluster Relaxation in Supercooled Chalcogenide Alloy Liquids

Given the configurational complexity of glasses and supercooled liquids, one might well have supposed *a priori* that $\beta$ would have no microscopic significance, and would vary inexplicably from one material to another, and from one probe to the next, even for the same material. Surprisingly enough, this view is still widely held [30], in spite of the numerous successes of the universal minimalist SLP model, both systematic [1] and predictive [2]. Those successes, however, are not readily achieved for data from poorly prepared samples (for example, partially crystallized polymers, where relaxation can be dominated not by intrinsic diffusion, but by nucleation kinetics).

Diffraction is a technique which at first seems to raise special problems, as $\beta = \beta(q)$, and one can expect that $\beta(q) \to 0$ as $q \to 0$ (low probe energy), while $\beta(q) \to 1$ for $q \gg Q_1$. Experiments on molecular glasses (OTP and other examples listed in Table I), as well as several numerical simulations, all agree that $\beta(Q_1)$ [the largest peak in the scattering function $S(Q)$ occurs at $Q = Q_1$] is in excellent agreement with non-diffraction probes [specific heat, untrasonics, light scattering], so that the choice $Q = Q_1$ is remarkably successful in converting the scattering data to quasi-particle kinetics.

It is clear that a different meaning must be attached to relaxation in supercooled liquids of excitations associated with the Boson (or first sharp diffraction) peak; if Kohlrausch relaxation was the oldest (150 years) unsolved problem in science, then explaining $\beta(Q_B)$ must be the most difficult problem for theory (or at least condensed matter theory). Here we discuss the first data on $\beta(Q_B,T)$ obtained by scattering spin-polarized neutrons for six binary chalcogen melts [31]. The relaxation here refers to some kind of cluster, and one might well suppose that here (at least) theory will finally fail. The present theory does not succeed in predicting the data to a few % accuracy (as it did for $\beta(Q_1)$), but it does explain systematic chemical trends. Moreover, these unique experiments test the significance of the effectiveness factor f in d* = fd in a wholly unexpected way. Historically such unique experiments can provide strong support for heuristic or axiomatic model theories.

Boson peaks, either in diffraction, Raman scattering, or vibrational spectra (either theoretical or from neutron scattering) are the characteristic signature of extended glass clusters [32]. In molecular metallic, or colloidal glasses (such as orthoterphenyl, OTP) boson peaks are associated with cages around vacancies or miscoordinated (relative to the crystal) sites [33,34], while the boson peak in metallic glasses is modeled elastically in terms of local structural shear rearrangements [35].

Chalcogenide network glass alloys have traditionally provided the best examples of FSDP and Boson peaks [32,36]. There the FSDP greatly increases in strength from Se chains to predominantly tetrahedral $GeSe_2$, where it corresponds to the Ge-Ge interplanar spacing. Detailed studies of the composition dependence of $Q_B$ and the amplitude of the FSDP show that it is weak in Se and strong in the layered compound compositions $As_2Se_3$ and $(Si,Ge)Se_2$. At intermediate cation concentrations there is an abrupt crossover from Se inter-chain dominated spacings to interplanar spacings for cation concentrations near 0.1, where the height of the FSDP also shows an abrupt change in slope [37].

Chalcogenide alloy glassy networks exhibit a Boson-like peak even in their vibrational densities of states measured by neutron scattering, called "floppy modes" [38]. With increased cation cross-linking the network stiffens, and in mean-field theory (constraint theory) this gives a stiffness transition [3,39]. Well-homogenized glasses self-organize and form percolative networks near the stiffness transition, giving rise to an exponentially complex intermediate phase (IP) [40]. Internal network stress is reduced by a large factor in the IP, as measured by Brillouin scattering for both chalocogenide alloys [41] and sodium silicate alloys [42]. Most of the rigid modes have condensed to form percolative paths [39,43,44], but some may have condensed to form small clusters which contribute to the FSDP.

The structural effects of internal network stress observable by diffraction are small, but they were evident in the reduced width of the FSDP in the average over 4 closely spaced compositions in the $Ge_xSe_{1-x}$ IP window $0.20 \leq x \leq 0.25$, compared to 14 closely spaced compositions outside the window (Fig. 9 of [45], reproduced and annotated here as Fig. 3). Moreover, as pointed out by Prof. G. Lucovsky (private communication), the PDF width shows a sharp second minimum near $x = 0.30$, where Raman spectra (1977) had show that a stress-relieving ethane-like structure $(Se_{1/2})_3$-Ge-Ge-$(Se_{1/2})_3$ is present [46,47]. There is an interesting technical point here. To compare EXAFS width data, which contain subcomponents of the PDF data, with the PDF width data for the FSDP, the authors of [45] combined the former assuming that there were no Ge-Ge contacts in the network. As can be seen from Fig. 2, this had the effect of erasing the sharp minimum at $x = 0.30$. Thus even had the Raman data not been known, in principle careful study of the differences between the PDF and combined EXAFS data would have suggested the onset of the ethane-like structure $(Se_{1/2})_3$-Ge-Ge-$(Se_{1/2})_3$ near $x = 0.30$. Moreover, since the effect shows up much more strongly in the FSDP than in Raman scattering, one can conclude that the ethane-like structure dominates the FSDP cluster for x above $x = 0.30$.

While the interpretation of their data given by the authors of [45] was overly conservative, it is true that diffraction is not the only tool for studying the FSDP clusters. With spin-polarized neutron scattering one can study the relaxation of clusters near $Q = Q_B$ in chalcogenide (Se,As, Ge) alloy melts [31]. The data here are complex and involve high temperatures (the lowest temperatures studied range from 1.2 $T_g$ to 1.6 $T_g$), where the clusters are much weaker than near $T_g$. However, the data contain some interesting structural features and one spectacular trend.

First it is helpful to picture the effects of adding cross-linking As or Ge to Se chains. The difference between As and Ge cross-linking is simply that As will be asymmetrically bonded (twice to one chain, once to the other), whereas the Ge bonding will be symmetrical (twice to each chain). Thus the Ge cross-linking clusters preserve the equivalence of the Se chains, and the activation energies $E_a(<r>)$, where $<r>$ is the average coordination number in the alloys, are a linear function of <r> from <r> = 2.0 (Se) up to <r> = 2.4 in $GeSe_4$, but the line drawn through the As alloys does not pass through Se (see Fig. 3 of [31]). (The same feature is evident in the temperature coefficient $C_\beta$ defined below.) This means that the relaxation of cross-linking clusters is more complex in the As alloys than the Ge alloys, which will reappear in chemical trends in $\beta(Q_B)$.

The most striking feature of the data of [31] is their discovery that $\beta(Q_B) = \exp(-C_B(x)/T)$, where x denotes the composition of the alloy. In the SLP relation $d^* = fd$, we expect the effective fraction f of relaxation channels to be small for large clusters (most degrees of freedom will not be effective, which again helps to explain why the chemical trends in are more complex for the As than the Ge alloys). With $f_B \ll 1$, $\beta(Q_B) = f/(2/d + f)$ is nearly proportional to f. One would expect the effective fraction f of relaxation channels to be thermally activated, which is exactly what [31] discovered. This is a totally unexpected way of confirming the physical meaning of f, and actually proving that f is an effective fraction.

The chemical trends in $C_B(x)$ are informative. Because Se chains in the melt are long [1], $C_B(x)$ should be smallest in Se, and increase with <r>, as observed. In fact, the reported values of $\beta_B(Se)$ are close to 0.43 near $T_g$, which is nearly the same value as $\beta(Q_1)$ in Se (Table I). In other words, for Se clustering effects are small, and the relaxation of the long chains is nearly independent of Q. However, as soon as cross-linking begins ($Ge_{08}Se_{92}$ in inset of Fig. 4 of [31]), β and f drop sharply (by about 1/3). Now one no longer has merely competition between short- and long-range forces (as in Se), but direct competition between effective and ineffective relaxation modes of clusters.

At these large values of $T/T_g$ the separation between $Q_1$ and $Q_B$ relaxation channels is not sharply defined in β(Q), but it is still perceptible where the clusters are best defined and f is smallest ($AsSe_3$ in Fig. 2 of [31] near Q = 1.8$A^{-1}$). In their unpublished data (private communication) for $GeSe_4$ near 1.55$A^{-1}$ there is a break in slope suggestive of incipient phase separation at 720K (~ 1.5 $T_g$ ). For the same value of <r> f is smaller for the As than for the Ge alloys, consistent with the activation energy differences discussed above.

## 6. Conclusions

We have seen that the SLP model is capable of drawing extremely detailed conclusions about SER for a wide variety of materials; indeed, each of the three examples discussed here is representative of the state of the art for its own class, and each could well stake a claim to being "best ever' in terms of providing rich insights into the complex relaxation processes first defined more than a century ago by Arrhenius. Perhaps most satisfying is the progress that these experiments have made possible in identifying microscopic configuration coordinates in the exponentially complex context of glasses and deeply supercooled liquids. Far from being a limitation [20] of the SLP theory, microscopic configuration coordinates are the substantive basis for connecting the abstract mathematical (asymptotic) aspects of SER to real experiments, as has been shown here for three elegant examples.

*Postscript.* After this paper was completed, another very interesting example of SER appeared, NMR spin-lattice relaxation and stimulated echoes in ice II, which behaves quite differently from ice I [48]. (This in itself is surprising, as ice I is hcp, and ice II is ccp, an apparently small difference.) The SER of defects in ice I showed large β ~ 0.8-1.0, while that of defects in the high-pressure phase ice II have β = 0.6, a difference the authors found puzzling. Comparing these data with Fig. 2 of this paper, one is led to a plausible model. The defects in as-grown ice I are large-scale relative to a single molecule (volumes about three times the host volume), whereas those in high-pressure ice have nearly the same volume, the "point" volumes presumably caused by high pressure internal stresses. The special value β = 1/3 (d* = 1) for spin glasses is now well-supported by at least two experiments [49]. Also self-bleaching in photodarkened Ge-As-S amorphous thin films exhibits SER with β = 3/7 (not surprising, since photodarkening certainly involves competition between long wave length darkening and molecular scale rebonding) [50], so that the topological list in Table I is by no means exhaustive.

| Method | Material | β(exp) | β(theory) | d* |
|---|---|---|---|---|
| Num. Simula. | Spin glass | 0.35 | 0.33 | 1 |
| Num. Simula. | Binary soft sphere | 0.62 | 0.60 | 3 |
| Num. Simula. | Coord. Alloy | 0.59 | 0.60 | 3 |
| Num. Simula. | Axial quasiX | 0.47 | 0.473 | 9/5 |
| Num. Simula. | Polymer | 0.59 | 0.60 | 3 |
| Stress Relax. | Se | 0.43 | 0.43 | 3/2 |
| Spin-Pol. Neutron | Se | 0.42 | 0.42 | 3/2 |
| Stress Relax. | Se-As-Ge | 0.61 | 0.60 | 3 |
| Stress Relax. | $B_2O_3$ | 0.60 | 0.60 | 3 |
| Stress Relax. | $Na_2O \cdot 4SiO_2$ | 0.63 | 0.60 | 3 |
| Stress Relax. | PVAC | 0.43 | 0.43 | 3/2 |
| Stress Relax. | PMA | 0.41 | 0.43 | 3/2 |
| Spin-Pol. Neutron | PB | 0.43 | 0.43 | 3/2 |
| Spin-Pol. Neutron | PVME | 0.44 | 0.43 | 3/2 |
| Spin-Pol. Neutron | PH | 0.44 | 0.43 | 3/2 |
| Spin-Pol. Neutron | KCN | 0.58 | 0.60 | 3 |
| Ultrasonic | KCN | 0.40 | 0.43 | 3/2 |
| Brillouin | KCN | 0.47 | 0.43 | 3/2 |
| Specific Heat | PG | 0.61 | 0.60 | 3 |
| Specific Heat | Glycerol | 0.65 | 0.60 | 3 |
| Specific Heat | OTP | 0.60 | 0.60 | 3 |
| Brillouin | OTP | 0.43 | 0.43 | 3/2 |
| Multidim NMR | OTP | (0.59,0.42) | (0.60,0.43) | (3,3/2) |

20| Method | Material | β(exp) | β(theory) | d* |
|---|---|---|---|---|
| Specific Heat | Salol | 0.60 | 0.60 | 3 |
| Ultrasonic | Glycerol | 0.60 | 0.60 | 3 |
| Brillouin | BD | 0.58 | 0.60 | 3 |
| Brillouin | HT | 0.60 | 0.60 | 3 |
| Brillouin | Salol | 0.60 | 0.60 | 3 |
| Spin-Pol. Neutron | OTP | 0.62 | 0.60 | 3 |
| Photo-Ind Absorp | $C_{60}$ | 0.40 | 0.43 | 3/2 |
| Photo-Ind Absorp | Cd(S,Se) | 0.40 | 0.43 | 3/2 |
| Carrier Relaxa. | a-Si:H | 0.44 | 0.43 | 3/2 |
| Carrier Relaxa. | porous Si | 0.4 | 0.43 | 3/2 |
| Photon Correl. | PG | 0.61 | 0.60 | 3 |
| Photon Correl. | PPG | 0.43 | 0.43 | 3/2 |
| Stress Relax. | PPG | 0.42 | 0.43 | 3/2 |
| Luminescence | Si nanodots | 0.57 | 0.60 | 3 |

Table I. Summary of values of stretching fraction β and topological dimensionality d* below $T_g$ previously discussed at length in [1,2]. There are 38 examples here. The present paper adds four more examples for Zn(Se,Te) alloys and OTP, as well as a discussion of the FSDP clusters in supercooled chalcogenide alloy liquids, and several more examples in the postscript, bringing the total close to fifty.

## Figure Captions

Fig. 1. Annotated data [23] on luminescence in isoelectronic Zn(Se,Te) alloys. The peak in relaxation time $\tau$ occurs near x = 0.10, where there is a break in $d\beta/dx$. These are the maximally localized states, which behave as quasi-particles subject to short-range forces only, with $\beta = 3/5$. The localized-extended transition occurs at x = 0.22, where the long- and short-range forces are equally weighted, and $\beta = 3/7$.

Fig.2. Relaxation in OTP [26] exhibits two branches of $\beta$, ad discussed in the text, corresponding to $d^* = 1$ and $d^* = 3+$.

Fig. 3. The measured widths of the FSDP are represented in this annotated figure from [31] by solid (black) circles; the other symbols refer to EXAFS data, which have been deconvoluted neglecting Ge-Ge contacts. Note that the measured widths of the FSDP exhibit structural features discussed in the text that are absent from the EXAFS data processed by neglecting Ge-Ge contacts.

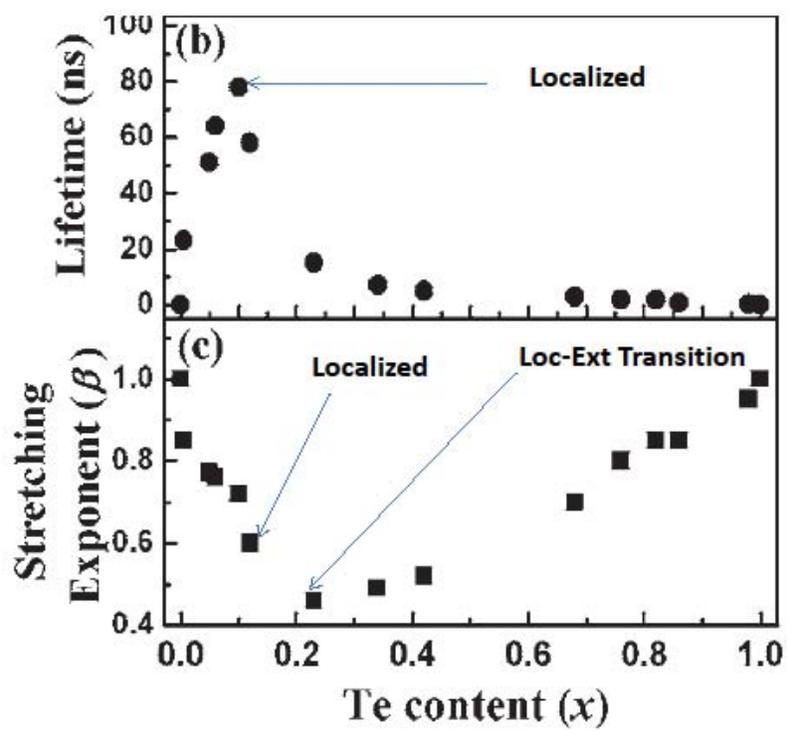

Fig. 1

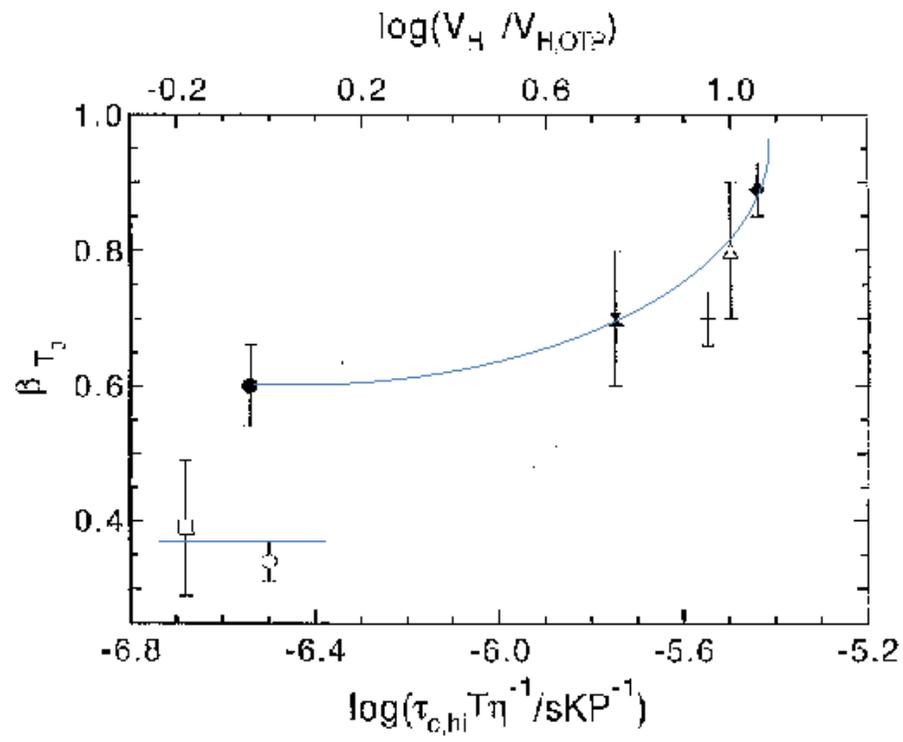

Fig. 2.

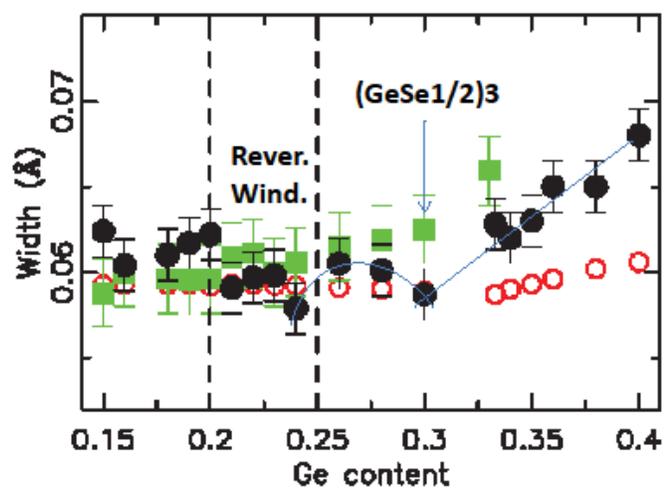

Fig. 3.